\documentclass[twocolumn,tighten]{aastex701}
\usepackage{amsmath}
\usepackage[utf8]{inputenc}

\newcommand{\customref}[2]{\hyperref[#2]{#1}}

\defcitealias{limbach_tempo_2023}{TEMPO I}
\defcitealias{soares-furtado_tempo_2024}{TEMPO II}

\begin{document}

\title{Twinkle Twinkle Little Star, Roman Sees Where You Are: 
Predicting Exoplanet Transit Yields in the Rosette Nebula with the Nancy Grace Roman Space Telescope}

\shorttitle{Predicting Exoplanet Transit Yields in the Rosette Nebula}
\shortauthors{Narayan et al.}

\author[orcid=0009-0007-0488-5685]{Ritvik Sai Narayan}
\affiliation{Department of Astronomy,  University of Wisconsin--Madison, 475 N.~Charter St., Madison, WI 53706, USA}
\affiliation{Wisconsin Center for Origins Research, University of Wisconsin--Madison, 475 N Charter St, Madison, WI 53706, USA}
\email[show]{rnarayan4@wisc.edu}  

\author[orcid=0000-0001-7493-7419,sname='Soares-Furtado']{Melinda Soares-Furtado}
\affiliation{Department of Astronomy,  University of Wisconsin--Madison, 475 N.~Charter St., Madison, WI 53706, USA}
\affiliation{Wisconsin Center for Origins Research, University of Wisconsin--Madison, 475 N Charter St, Madison, WI 53706, USA}
\affiliation{Department of Physics, University of Wisconsin--Madison, 1150 University Avenue, Madison, WI 53706, USA}
\email{mmsoares@wisc.edu}  

\author[orcid=0000-0002-9521-9798]{Mary Anne Limbach}
\affiliation{Department of Astronomy, University of Michigan, Ann Arbor, MI 48109, USA}
\email{mlimbach@umich.edu}

\author[orcid=0009-0004-7867-9479]{Nishanth Ramanujam}
\affiliation{Department of Astronomy, University of Michigan, Ann Arbor, MI 48109, USA}
\email{nramanu@umich.edu}

\author[0000-0001-7246-5438]{Andrew Vanderburg}
\affiliation{Department of Physics, Massachusetts Institute of Technology, 77 Massachusetts Avenue, Cambridge, MA 02139, USA}
\affiliation{Kavli Institute for Astrophysics and Space Research, Massachusetts Institute of Technology, Cambridge, MA 02139, USA}
\affiliation{Center for Astrophysics | Harvard and Smithsonian, 60 Garden Street, Cambridge, MA 02138, USA}
\email{avanderburg@cfa.harvard.edu}

\author[0000-0003-0489-1528]{Johanna M. Vos}
\affiliation{School of Physics, Trinity College Dublin, The University of Dublin, Dublin 2, Ireland}
\email{johanna.vos@tcd.ie}

\begin{abstract}
Young stars host only a small fraction of the known exoplanet population because their photometric variability, magnetic activity, and frequent placement in dense, poorly-resolved regions hamper exoplanet detections. Yet, measuring planets at these ages is crucial since these phases are when dynamical processes that drive planetary migration are most active. We assess the expected yield of a hypothetical Nancy Grace Roman Space Telescope transit survey of the Rosette Nebula, a ${\sim}10\,\mathrm{Myr}$ star-forming region with a dense and diverse stellar population. Using the Roman Exposure Time Calculator to quantify sensitivity to Rosette members, we establish detection thresholds for companions and evaluate yields via Monte Carlo injection-recovery simulations, accounting for nebular extinction and youth-driven stellar variability.
We predict the detection of $33 \pm 9$ young transiting exoplanets orbiting stellar hosts in a month-long survey, and $29 \pm 8$ in a two-week survey. The extended baseline primarily improves sensitivity to longer-period planets orbiting FGK stars, while most M dwarf detections are well-sampled within two weeks.
Irrespective of the temporal baseline, transit detections are dominated by of 1-2\,$R_\oplus$ super-Earths and sub-Neptunes with $P\lesssim8$\,days.
Such a sample would substantially expand the census of only three detected transiting planets younger than $20\,\mathrm{Myr}$ around stars less massive than the Sun, probing an age regime in which planetary radii remain inflated, the stability of close-in orbits is uncertain, and planetary migration may still be ongoing. This survey offers a path to constrain early planetary evolution and establish prime follow-up targets for JWST, Rubin, and the Habitable Worlds Observatory.
\end{abstract}


\section{Introduction} \label{section: intro}

Understanding how planetary systems form and evolve during their first few million years remains one of the central challenges in exoplanetary science. During this early epoch, planets are still contracting and cooling \citep{fortney_planetary_2007,linder_evolutionary_2019}, while their orbits and atmospheres are being actively sculpted by disk-driven migration, dynamical interactions, and stellar evolution \citep{kennedy_planet_2008,ida_toward_2008,cossou_hot_2014}. Detecting and characterizing young systems in this stage therefore provides a direct window into the physical mechanisms that determine planetary radii, compositions, and orbital architectures. These observations can help discriminate between in-situ and migration-dominated formation pathways \citep{mordasini_characterization_2012} and constrain the timescales for planets to reach thermal and dynamical equilibria \citep{owen_atmospheric_2019}. Furthermore, this assessment is also critical to understand planetary habitability, as the locations of habitable zones also evolve rapidly during the first few hundred million years \citep{kopparapu_habitable_2013}, influencing the composition of forming planets \citep{meadows_exoplanet_2018}.
Observing planets over this formative stage allows us to probe the progenitor population of mature, potentially habitable worlds. 

However, young transiting planets remain rare. There are only a small number of confirmed planets younger that $\lessapprox100$\,Myr \citep[e.g.,][]{mann_zodiacal_2017,david_four_2019,david_neptune-sized_2016,holczer_transit_2016,barber_giant_2024}. 
Observations of planets at distinct formative stages are necessary to trace the evolutionary pathways of these systems. Nevertheless, strong optical extinction and crowded environments in young star-forming regions (SFRs) make such observations extremely challenging with current ground-based and space observatories. 

The Nancy Grace Roman Space Telescope (hereafter Roman), NASA's next flagship astrophysics mission, is designed to deliver wide-field imaging and spectroscopy and provides an unprecedented opportunity to overcome these limitations. Scheduled to launch in late 2026, it will carry the Wide Field Instrument (WFI), a 300 megapixel near-infrared camera comprised of 18 detectors, and a $0.28\,\mathrm{deg}^2$ field of view (FOV), approximately 100 times larger than the Hubble Space Telescope's FOV.  Observations at these near-infrared wavelengths are less affected by dust extinction relative to optical bands ($A_{\mathrm{J}}/A_{\mathrm{V}}= 0.282$, $A_{\mathrm{H}}/A_{\mathrm{V}} = 0.175$, and $A_{\mathrm{K}}/A_{\mathrm{V}} = 0.112$), enabling a more complete census of reddened and embedded sources. Furthermore, unlike previous optical wide-field transit surveys such as \textit{Kepler}, \textit{TESS}, \textit{HATNet}, and \textit{WASP}, which become incomplete for low-mass hosts ($\lesssim 0.2\; M_\odot$) due to limited sensitivity at faint optical magnitudes \citep{borucki_kepler_2010,ricker_transiting_2015,bakos_hatnet_2018,smith_impact_2006}, observing in the near-infrared allows detections around cooler and redder low-mass stars. 

Several recent studies have demonstrated Roman's potential for exoplanet discovery across distinct Galactic environments: the Transiting Exosatellites, Moons, and Planets in Orion (TEMPO) survey (\citetalias{limbach_tempo_2023}: \citealt{limbach_tempo_2023}; \citetalias{soares-furtado_tempo_2024}: \citealt{soares-furtado_tempo_2024}) established its capability to detect and characterize transiting planets, satellites, and exomoons in the Orion Nebula Cluster (ONC) with a month-long survey; \citet{wilson_apache_2019} estimate that Roman's \textit{Galactic Bulge Time-Domain Survey} could detect up to ${\sim}10^5$ transiting exoplanets; \citet{terry_predictions_2025} constrain exoplanet detections with microlensing events; and \citet{kerins_rosetz_2023} outline the proposed \textit{Roman Survey of the Earth Transit Zone} (RoSETZ) to search for habitable-zone planets around nearby stars. Collectively, these efforts illustrate Roman's transformative capability to map the architectures of planetary systems across a wide range of stellar ages and environments. 

Building on this foundation, we outline a potential investigation of the Rosette Nebula, a dense SFR with a rich population of slightly older and more massive stars (Section~\ref{fig:isochrone}), in comparison to the ONC. This complementary environment offers a unique laboratory for investigating the formation pathways of planetary systems during the late disk-dissipation phase \citep{gaidos_diversification_2025}.

Beyond exoplanets, surveys of young SFRs enable a wide range of auxiliary science, including studies of stellar variability and activity in young stars, accretion and disk evolution, and potential transient phenomena, as outlined in \citetalias{soares-furtado_tempo_2024}. Together, these investigations highlight Roman's unique position to expand the current census of young exoplanets and provide the stellar and environmental context required to accurately interpret their formation and evolutionary pathways.

This paper is organized as follows: Section~\ref{sec: survey_design} details our survey design elements such as our target field of the Rosette Nebula, FOV orientation, and filter selection. In Section~\ref{section: stellar_population}, we present membership analysis and determination of this region's (sub)stellar population. We present the expected signal-to-noise for objects in the Rosette Nebula in Section~\ref{sec: exoplanet_detection_limit}. In Section~\ref{sec: simulation}, we introduce and discuss our modeling methodology to determine exoplanet transit yields. In Section~\ref{sec: summary}, we provide a brief summary of this work. Lastly, in the \customref{Appendix}{sec:appendix} we present Kernel Density Estimates (KDEs) of our recovered planet population from these simulations.

\section{Observational Setup} \label{sec: survey_design}

The survey outlined in this work is designed to achieve the primary science goal of detecting young transiting exoplanets in a previously unexplored region of parameter space: young, low-mass stellar hosts with small exoplanetary companions. In this section, we present the observational setup for Roman's wide field instrument (WFI) used to conduct this survey. 

\subsection{Target Field Selection} \label{subsec: target_field}

\begin{figure*}[t!]
    \centering
    \includegraphics[scale=0.65]{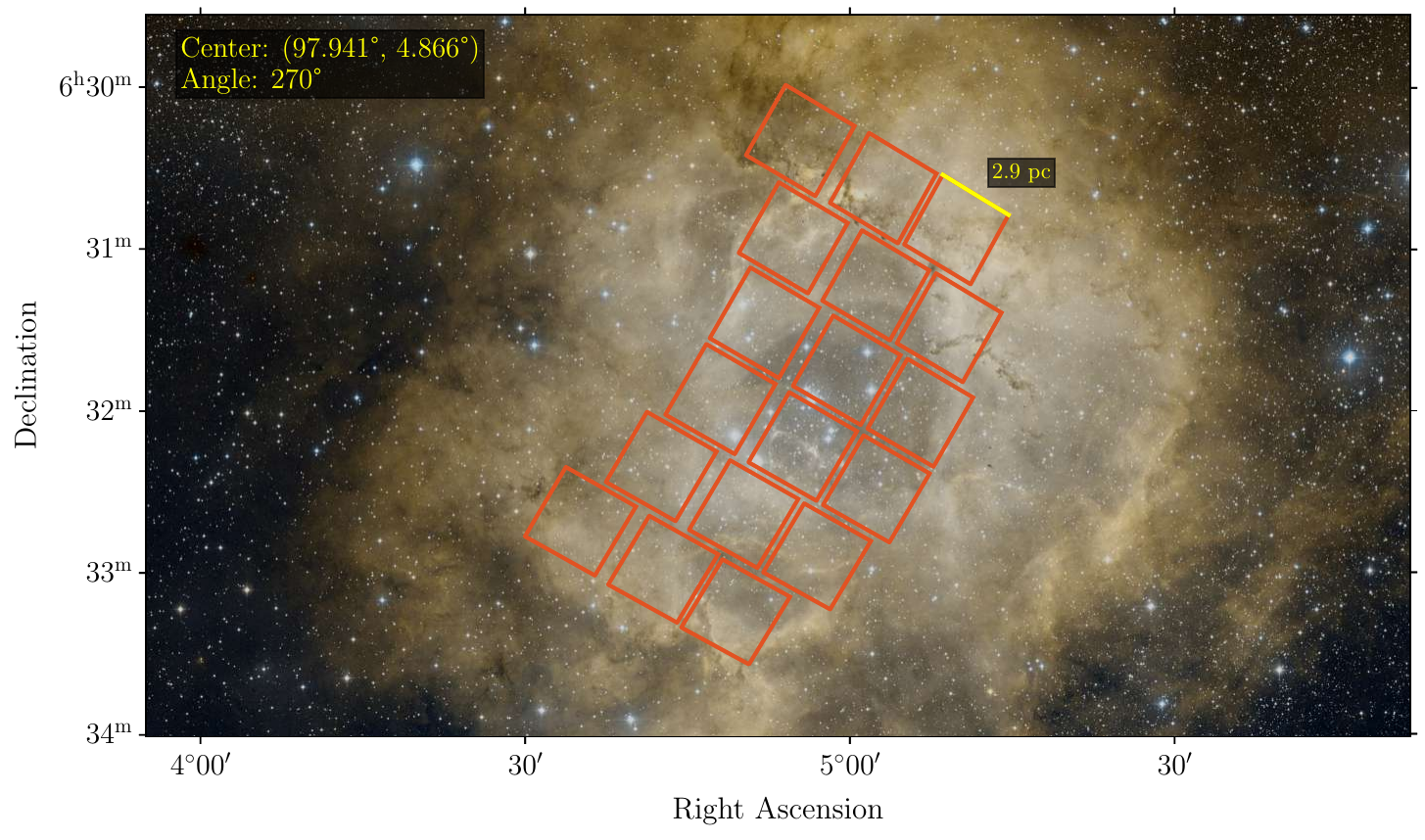}
    \caption{Roman WFI footprint overlaid on an image of the Rosette Nebula taken from the Second Digitized Sky Survey \citep[DSS-2;][]{lasker_guide_1990}. Red squares outline the 18 individual WFI detectors. The physical scale bar assumes a distance of 1.4\,kpc to the Rosette Nebula.}    \label{fig:footprint}
\end{figure*}

There exists a large population of young and dense SFRs in the Milky Way, including the ONC \citep[1--3\,Myr;][]{jeffries_using_2007}, Upper Scorpius, Lagoon Nebula \citep[1--2\,Myr;][]{venuti_multicolor_2021}, 30\,Doradus \citep[1--2\,Myr;][]{sun_deep_2022}, the Rosette Nebula \citep[0.5--10\,Myr;][]{muzic_looking_2019}, and Westerhout\,40 \citep[$< 7$\,Myr;][]{shuping_spectral_2012}. Among these, the Rosette Nebula stands out as an especially compelling target for a Roman exoplanet transit survey because (1) the region hosts thousands of young stellar objects spanning a wide range of spectral types (O5--L2) \citep[e.g.,][]{cambresy_young_2013,muzic_looking_2019}, (2) its distance (1400 pc) places members in an ideal regime across a range of (sub)stellar masses that is bright enough for high S/N yet faint enough to avoid saturation, and (3) the region comfortably fits within Roman's FOV as shown in Figure~\ref{fig:footprint}, with an estimated peak surface density of $\Sigma = 300 \mathrm{\; stars \; pc^{-2}}$ \citep{kuhn_spatial_2014, kuhn_spatial_2015}. 

At the system's distance of approximately 1400\,pc \citep{muzic_looking_2019}, FGK-type stars remain below the Roman Wide Field Instrument's saturation limit, while high photometric precision (Table~\ref{tab:mags}) is preserved for low-mass stars and brown dwarfs (BDs). Moreover, \textit{Gaia} \citep{gaia_collaboration_gaia_2016} astrometric measurements reveal that the stars in the Rosette Nebula occupy a single, well-defined locus (see middle panel in Figure~\ref{fig:membership}), indicating largely aligned kinematic motions. This property simplifies membership analysis of the region in comparison to kinematically complex regions with multiple velocity components such as Westerhout~40 \citep{shimoikura_dense_2015}. 

We summarize the Rosette survey parameters in Table~\ref{tab:survey_overview}. 

\begin{deluxetable}{lr}[htb!]
\tablewidth{0pt} 
\tabletypesize{\scriptsize}
\tablecaption{Observational Parameters for the Rosette survey, which probes the Rosette Nebula and neighboring regions. The F146 spectral band was used in the calculation of the magnitude limit and photometric precision. \label{tab:survey_overview}}
\tablehead{\colhead{Parameter} & \colhead{Value}}
\startdata
\multicolumn{2}{c}{\textit{Observational Parameters}} \\
\hline
Field of View & 0.28\,deg$^2$ \\
Spectral Band & F146 (0.93--2.00\,$\mu$m) \\
Exposure Time & 6\,reads, 18\,sec \\
Photometric Precision (1 hr) &  125\,ppm (17\,mag$_\mathrm{AB}$) \\
                             & 850\,ppm (21\,mag$_\mathrm{AB}$)\\
Duration & 14 days \\
         & 30 days \\
Magnitude Limit & 29.3\,mag$_\mathrm{AB}$ (14-day)\\
 & 29.7\,mag$_\mathrm{AB}$ (30-day) \\
\hline
\multicolumn{2}{c}{\textit{Estimated Number of Monitored Sources}} \\
\hline
Young FGK stars & 409 \\
Young M stars & 1290 \\
Young Brown Dwarfs & 1125 \\
\enddata
\end{deluxetable}

\subsection{Field of View \& Orientation} \label{subsec: fov}

We defined the survey footprint by combining the \textit{Gaia} DR3 catalog, including sources down to a faintness limit of $G \approx 21$\,mag \citep{vallenari_gaia_2023}, with the Roman WFI footprint available through the Aladin Lite v3 interactive sky atlas \citep{baumann_aladin_2022}.\footnote{\url{https://aladin.cds.unistra.fr/AladinLite/}} For this analysis, we considered sources falling within the boundaries of the 18 WFI detectors, omitting dithers. 
To explore the impact of orientation, we sampled 12 different position angles in 30$^\circ$ increments. The field center was set to coincide with the densest stellar region at the heart of the Rosette Nebula, located at $(\alpha, \delta) = (06^\mathrm{h}\,31^\mathrm{m}\,45.81^\mathrm{s}, +04^\circ\,51'\,58.0'')$, with an orientation angle of $270^\circ$ selected to maximize the number of monitored stars. In this orientation, the Roman footprint encompasses a total of 11,352 stars from the \textit{Gaia} DR3 catalog. We use this as our precursory sample to determine the stellar population that Roman will monitor in the Rosette Nebula (as discussed in Section~\ref{section: stellar_population}). 

\subsection{Filter Selection} \label{subsec: filter}

To optimize sensitivity to the detection of exoplanets transiting low-mass host stars, we employ the Roman/WFI broadest near-infrared F146 filter ($0.9$--$2.07\,\mu$m; pivot wavelength 1.44\,$\mu$m). This bandpass encompasses the $Y$, $J$, and $H$ windows as well as the gaps between them, providing high throughput and excellent sensitivity to cool stars whose spectral energy distributions peak in the near-infrared. The wide spectral coverage yields strong signal-to-noise for faint M dwarfs, the dominant stellar population in coeval populations. We also consider the use of the F213 filter ($1.88$--$2.38\,\mu$m; pivot wavelength 2.12\,$\mu$m), which offers $K$ band coverage optimized for cool, late-type hosts, whose longer wavelengths may also reduce the impact of stellar spot contrast \citep{schutte_measuring_2023}. While narrower WFI filters such as F213 may assist in stellar characterization, F146 allows us to probe faint sources at high signal-to-noise.

\subsection{Exposure Time and Cadence Selection} \label{subsec: cadence}

Similar to the exposure time and cadence selection protocol of the TEMPO survey, adopting 6-read integrations with individual exposure times of 18\,seconds. This configuration ensures the minimum number of integrations recommended for both high-precision photometry and effective cosmic-ray mitigation, while ensuring that we preserve the dynamic high signal-to-noise across bright and faint stellar populations. We also note that our resulting cadence is sufficient to temporally resolve ingress and egress, including close-in planets whose shortest simulated transit (Section \ref{sec: simulation}) lasts ${\sim} 2$\,hr. 

\section{Assessing the (Sub)Stellar Population of the Rosette Nebula} \label{section: stellar_population}

Having established the observational setup, we now quantify the number and types of sources the WFI will observe in the proposed Rosette survey, with emphasis on cluster members expected within the WFI footprint given our adopted exposure times, cadence, and magnitude limits. With our \textit{Gaia} sample, we are unable to detect most low-mass stars that Roman will monitor, and hence we extrapolate the initial mass function (IMF) to estimate the (sub)stellar population.

\subsection{Membership Determination} \label{subsec: membership}

Kinematics are often key in distinguishing member stars from those in the field. Common clustering methods, such as $k$-means or Gaussian mixture models, rely on astrometric data but require assumptions about cluster number and shape, and they lack robust mechanisms for handling noise or outliers. To overcome these limitations, we combine unsupervised machine learning methods with well-established system priors to build a robust sample of members.

\begin{figure*}[htb!]
    \centering
    \includegraphics[width=\linewidth]{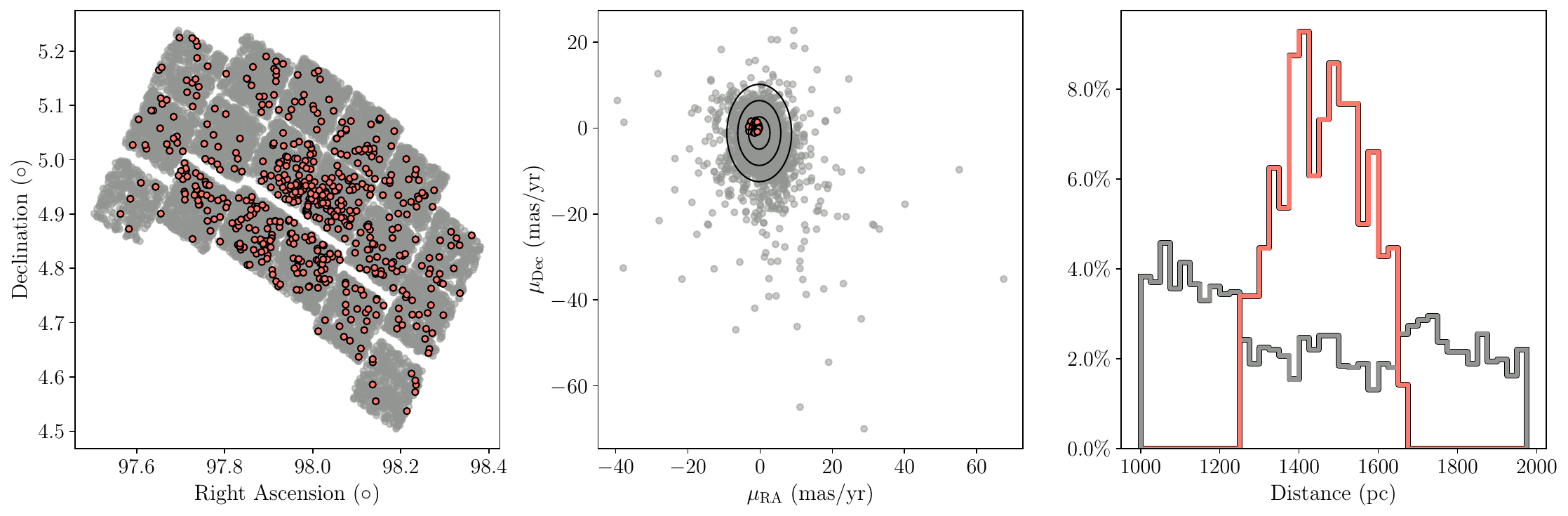}
    \caption{Gaia DR3 sources overlapping the Roman/WFI footprint in the Rosette Nebula. Candidate members (salmon) and field stars (gray) are shown in their sky distribution (left), proper motions (center), and distances (right). We also show $1\sigma$, $2\sigma$, and $3\sigma$ proper-motion contour lines for field stars (center). To reduce foreground and background contamination, we applied a distance prior restricting the sample to sources within 1–1.9\,kpc.}
    \label{fig:membership}
\end{figure*}

We began with the 11,352 \textit{Gaia} DR3 sources that overlap the Roman/WFI footprint shown as a combination of gray and salmon-hued points in the left panel of Figure~\ref{fig:membership}. To reduce foreground and background contamination, we applied a distance prior restricting sources to 1–1.9\,kpc. This range was guided by previous distance estimates for the Rosette Nebula \citep[e.g.,][]{muzic_looking_2019,muzic_stellar_2022}, but expanded to be slightly more generous in order to allow for the possibility of additional members at larger separations.

To identify a core member of the Rosette Nebula that could serve as a kinematic reference to find comoving member stars, we applied Hierarchical Density-Based Spatial Clustering of Applications with Noise \citep[HDBSCAN;][]{campello_density-based_2013} to five parameters (RA, Dec, proper-motion in RA ($\mu_{\alpha}$), proper-motion in Dec ($\mu_{\delta}$), and parallax ($\pi$)) from this sample. The core member Gaia DR3~3131334834450613376 ($\overline{\mathrm{RV}} = 38.01 \, \mathrm{km \, s^{-1}}$, \citealt{almeida_eighteenth_2023}) was identified. The proper motion of this core member ($\mu_{\alpha} = -1.67 \;\mathrm{mas \, yr^{-1}}$, $\mu_{\delta} = 0.24 \;\mathrm{mas \, yr^{-1}}$) was in good agreement with the proper-motion distribution identified by  \citet{muzic_stellar_2022} in a prior analysis of the Rosette Nebula. 

To further refine our sample of likely Rosette Nebula members, we conducted a volume-limited (200\,pc) kinematic search using the \textsf{FriendFinder}\footnote{\url{https://github.com/adamkraus/Comove}} algorithm \citep{tofflemire_tess_2021} to identify stars whose tangential velocities were within $\pm 10 \; \mathrm{km \, s^{-1}}$ of the values predicted for their positions if they were comoving with our chosen core member. We adopted a comparatively generous $\pm 10 \; \mathrm{km \, s^{-1}}$ window to avoid excluding bonafide members given the non-virial, expanding kinematics of the Rosette Nebula, where the tangential velocity is expected to increase with radius from the core \citep{lim_kinematic_2021,muzic_stellar_2022}. 
We opted not to include radial velocities (RVs) as an additional kinematic constraint, as most stars in this highly-extincted region lack reliable RV measurements.

This procedure yielded 546 candidate Rosette Nebula members. As illustrated in Figure~\ref{fig:membership}, the candidate members are more concentrated toward the center of the Roman/WFI footprint, following the known Rosette Nebula structure, while the field stars are spread more uniformly across the field. Further, in proper-motion space, the Rosette Nebula members form a tight clump with mean values of $\bar{\mu}_{\alpha} = -1.58 \pm 0.5 \, \mathrm{mas \, yr^{-1}}$ and $\bar{\mu}_{\delta} = 0.23 \pm 0.5 \, \mathrm{mas \, yr^{-1}}$, which is distinct from the broader distribution of field stars and overlap the distributions from prior investigations \citep{muzic_stellar_2022}. In distance space, the Rosette Nebula members exhibit a strong peak centered at 1.4\,kpc (see Figure~\ref{fig:membership}), which we adopt as our distance to the Rosette Nebula in further analyses. Our adopted distance is consistent with literature estimates for the Rosette Nebula: 1.39\,kpc \citep{hensberge_eclipsing_2000}, 1.48\,kpc \citep{muzic_stellar_2022}, 1.59\,kpc \citep{muzic_looking_2019}, and 1.6\,kpc \citep{phelps_parsec-scale_2005}.

Together, the three panels demonstrate that the identified members are spatially concentrated, kinematically distinct, and distance-coherent, which strengthens confidence in our membership selection. Our Rosette Nebula sample extends to an angular radius of $0.42^\circ$, corresponding to ${\sim}$10.3\,pc at a distance of 1.4\,kpc.

\subsection{Isochrone Fit} \label{subsec: isochrone}

\begin{figure}[h!]
    \centering
    \includegraphics[width=\linewidth]{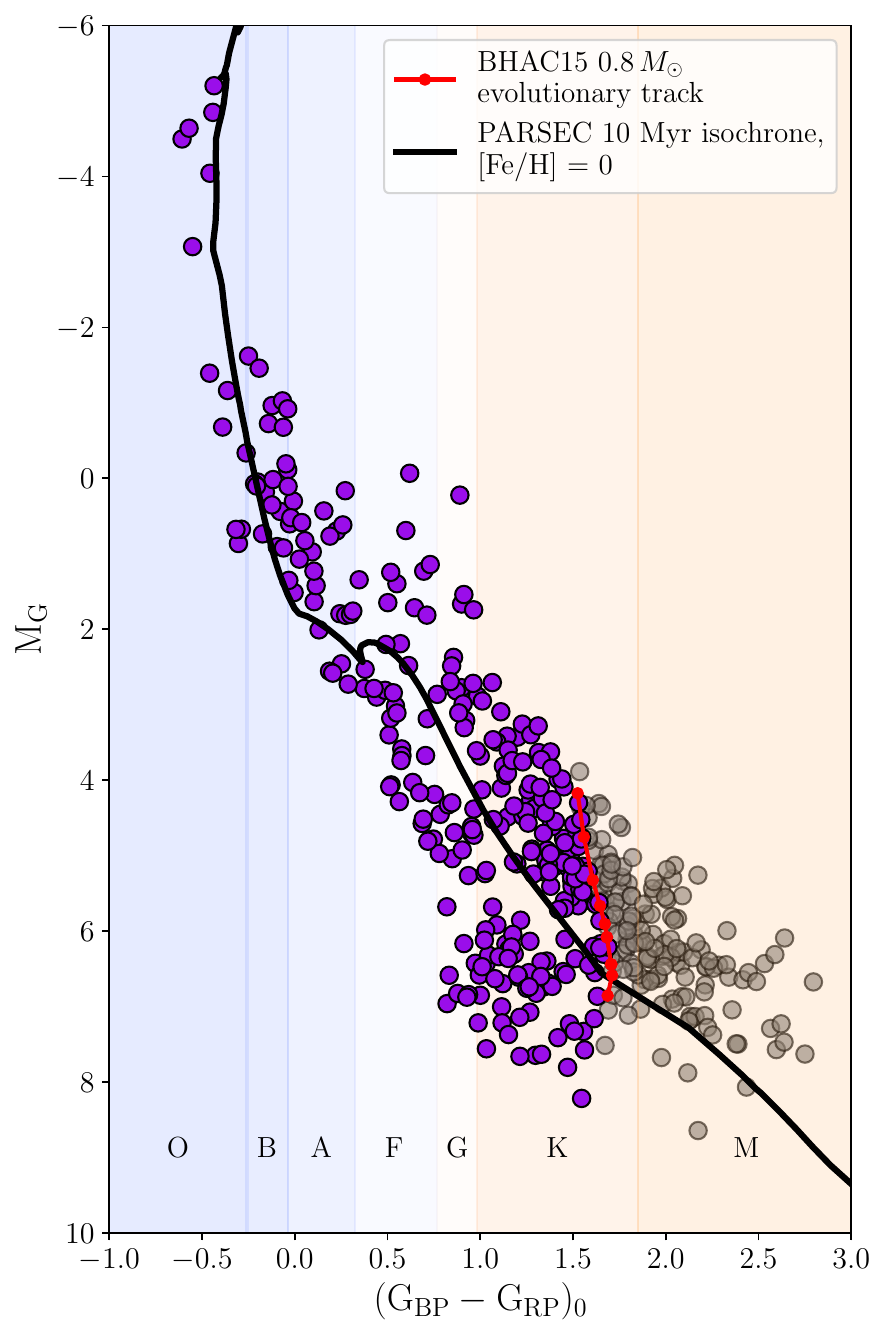}
    \caption{We show our extinction and reddening corrected \textit{Gaia} sample (circles) in absolute color-magnitude space and the \cite{baraffe_new_2015} evolutionary track for a $0.8 \: M_\odot$ star (red line) with points at 0.5\,Myr, 1\,Myr, 2\,Myr, 3\,Myr, 4\,Myr, 5\,Myr, 7\,Myr, 10\,Myr, and 20\,Myr (red circles). The purple circles represent stars with interpolated masses above our completeness limit of $0.8 \; M_\odot$, whereas the gray circles represent members below this limit that we discard in our determination of the initial mass function normalization. We also plot a 10\,Myr solar-metallicity \textsf{PARSEC} isochrone (black line) and show the spectral classifications corresponding to each color bin.}
    \label{fig:isochrone}
\end{figure}

As pre-main sequence stars converge onto the zero age main sequence, they exhibit substantial dispersion in their color-magnitude diagram positionality due to effects such as stellar variability, accretion, magnetic activity, and starspots \citep{bell_pre-main-sequence_2013,somers_older_2015}. Further, star formation does not happen instantaneously across a SFR, often being triggered sequentially from feedback from massive stars. This results in a genuine age distribution of a few Myr. \citet{muzic_stellar_2022} demonstrated that this was also true of the Rosette Nebula region (0.5-10\,Myr). In addition to a true dispersion in age, some of the scatter can arise from observational uncertainties, such as challenging extinction/reddening constraints and binarity.

To fit an isochrone to our membership sample, we first corrected for dust extinction and reddening. \textit{Gaia} DR3 provides \textsf{Apsis}-based extinction/reddening estimates for only ${\sim}eq 470$ million of nearly two billion sources, so many stars lack \textit{Gaia}-provided corrections \citep{creevey_gaia_2023,vallenari_gaia_2023}. To maximize the number of usable members in this analysis, we use the median $\mathrm{E(G_{BP} - G_{RP})}$ and $\mathrm{A_G}$ derived from a hierarchical binning approach for stars that lack individual \textsf{Apsis}-provided corrections. We first compute the uncorrected stellar color and absolute G-magnitude, dividing these into quantile-based bins. For each missing extinction value, we apply a three-tier strategy: (1) use the median from stars in the same 2D color-magnitude bin, (2) if insufficient data exists, use the median from stars with the same color only, or (3) as a final fallback use the global sample median.  

For our analysis, we adopt a single fiducial age of 10\,Myr at solar metallicity. We present a \textsf{PARSEC} isochrone \citep{bressan_parsec_2012,chen_improving_2014} with these parameters for our sample in Figure~\ref{fig:isochrone}. Given that star formation occurs in multiple epochs, choosing the upper limit on the Rosette Nebula's age allows us to consider more stars that have converged onto the main sequence. We find that the bluer stars ($\mathrm{(G_{BP} - G_{RP})_0} \lessapprox 0$) are best fit in color-magnitude space with the 10\,Myr isochrone. 

We assign masses to each star in our sample by interpolating their extinction-corrected CMD positions onto the 10\,Myr \textsf{PARSEC} isochrone. 
Although the intrinsic stellar mass function rises steeply toward lower masses, the observed number of stars drops below ${\sim} 0.8 \; M_\odot$, indicating incompleteness in our \textit{Gaia} sample in this low-mass regime. We therefore adopt $0.8 \: M_\odot$ as our completeness limit and restrict our initial mass function (IMF) normalization analysis to the 299 member stars (purple points in Figure~\ref{fig:isochrone}) brighter and bluer than the \citet{baraffe_new_2015} evolutionary track for a star of this mass at solar-metallicity. Stars fainter/redder than this limit (gray points) are excluded to avoid biases from incompleteness at lower masses, with the near-vertical morphology of the $0.8 \; M_\odot$ evolutionary track also providing a natural mass-partition. 

Stellar multiplicity exhibits a well-established positive correlation with stellar mass \citep{duchene_stellar_2013}, implying that binarity may bias IMF normalizations. We correct for these effects on our sample by accounting for multiplicity fractions from Table~1 in \citet{duchene_stellar_2013} and assume a flat mass-ratio distribution ($q {\sim} 0.5$). We adopt this simplification because our mass-completion cuts exclude M dwarfs, where $q \simeq 1$ \citep{henry_character_2024}, whereas most companions for OBA stars remain in our high-mass bin so the detailed mass-ratio distribution has a negligible effect on our final count. Finally, this also prevents overcounting for FGK stars \citep{raghavan_survey_2010}. This yields us a multiplicity-corrected sample size of 413 stars. 

\subsection{Initial Mass Function} \label{subsec: imf}

We use our final sample to construct the IMF of the Rosette Nebula. We adopt a broken power-law parameterization of the IMF of the form, 

\begin{equation}
    \frac{dN}{dm} = km^{-\alpha} 
\end{equation} \label{eqn: imf}

where $k$ is a normalization constant and $\alpha$ is the slope of the IMF. In defining the corresponding slopes, we use the following two regimes: 

\begin{itemize}
    \item For $0.8\,M_\odot\leq m \leq 20\, M_\odot$, we use a Salpeter slope of $\alpha = 2.35$ \citep{salpeter_luminosity_1955}. \citet{muzic_looking_2019} found that the stellar population in NGC\,2244 (the core of the Rosette Nebula) is well-described with a high-mass slope close to the Salpeter value. This mass range corresponds to the portion of our sample that is demonstrably complete, and thus use this bin to find the normalization constant, $k = 419$. We note that extremely massive stars ($m \gtrapprox 20 \, M_\odot$) contribute negligibly to this normalization since the Salpeter slope causes their relative numbers to decline steeply.
    \item For $0.01\,M_\odot\leq m < 0.8\, M_\odot$, we adopt $\alpha = 1.05$ as measured by \citet{muzic_stellar_2022}, who derive this slope by mapping observed luminosity functions to masses using \textsf{BT-Settl} evolutionary tracks \citep{baraffe_new_2015}, combined with \textsf{PARSEC} isochrones \citep{bressan_parsec_2012}, and bolometric corrections \citep{pecaut_intrinsic_2013,hewett_ukirt_2006}. This prescription provides a physically motivated, empirically tested description of the IMF down to planetary-mass objects.
\end{itemize}

Integrating Equation~\ref{eqn: imf} with the relevant mass bins, we estimate 409 FGK stars, 1290 M dwarfs, and 1125 BDs that we will monitor. We also show the estimated number of monitored BDs, M dwarfs, and FGK stars in Table~\ref{tab:survey_overview}.

\section{Determining the Exoplanet Detection Limit} \label{sec: exoplanet_detection_limit}

In this section, we present our methodology for modeling the signal-to-noise ratio (S/N) and detection magnitudes for free-floating planets (FFPs), BDs, and stars in the Rosette Nebula. We assess the detectability of transiting exoplanets orbiting young stellar hosts with masses between $0.08\mbox{-}0.9$\,$M_\odot$. We investigate transit yield constraints with both the F146 and F213 Roman filters in order to determine which of these filters would result in a higher transit yield. 

\subsection{Synthetic Photometry} \label{subsec: synthetic_photometry}

\begin{figure*}[hptb]
    \centering
    \includegraphics[scale=0.6]{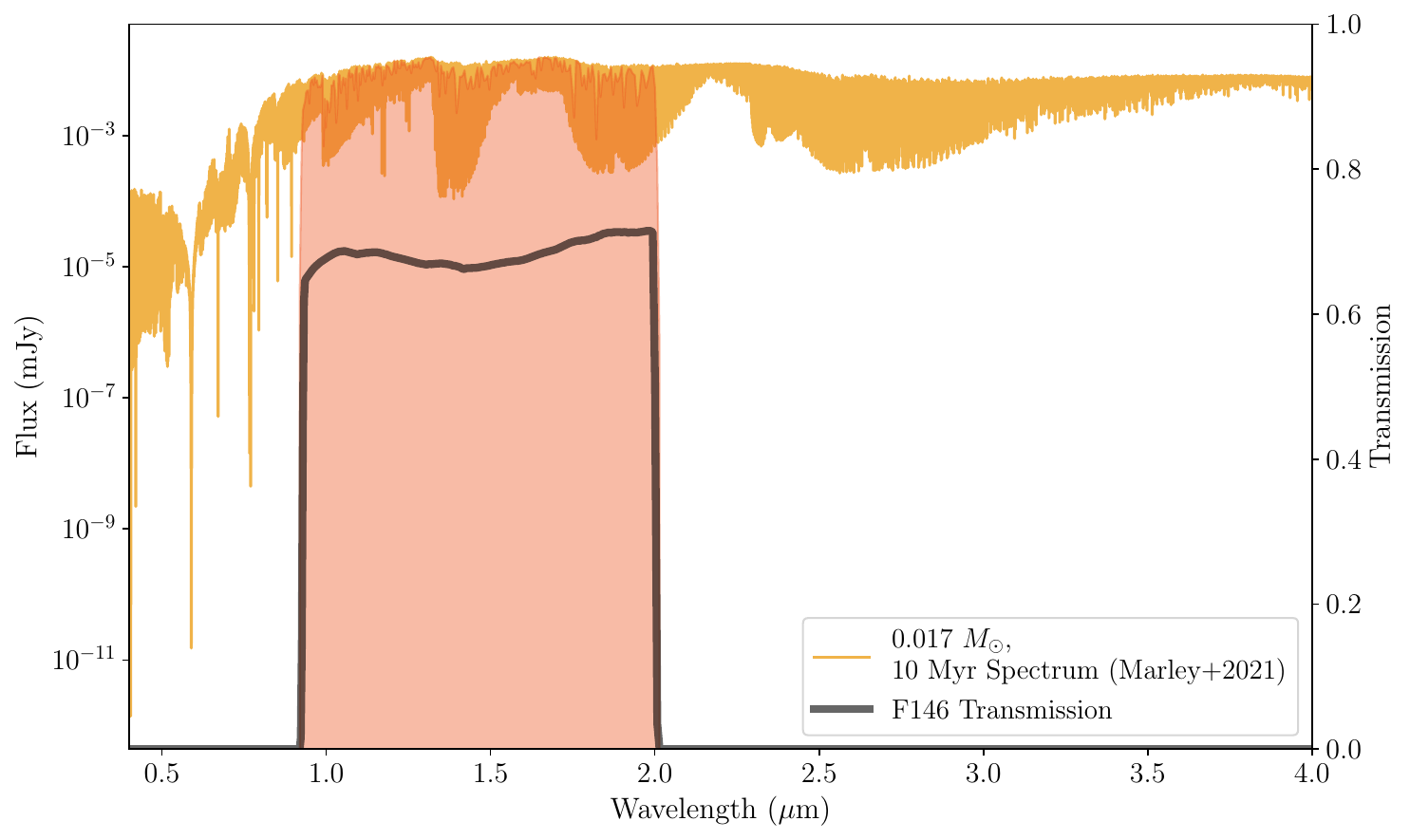}
    \caption{We present a 10\,Myr brown dwarf \citet{marley_sonora_2021} model spectrum (orange line). We also show the normalized transmission of the Roman broadband F146 filter (black line) and its convolution with the spectrum (orange shaded region), corresponding to its effective stimulus.}
    \label{fig:transmission}
\end{figure*}

To compute magnitudes of FFPs, BDs, and stars in the two Roman/WFI filters we combine, 

\begin{itemize}
    \item $0.002 \; M_\odot \leq m < 0.02  \; M_\odot$ ($2.1 \; M_{\rm J} \leq m < 21 \; M_{\rm J}$): \textsf{Sonora Bobcat} self-luminous sub-stellar atmospheres \citep{marley_sonora_2021} with age-dependent evolutionary models from \citet{saumon_evolution_2008}.     
    \item $0.02 \; M_\odot \leq m < 0.2  \; M_\odot$: \textsf{PHOENIX} model atmospheres \citep{hauschildt_parallel_1997,hauschildt_nextgen_1999}, with age-dependent evolutionary models from \citet{baraffe_new_2015}. 
    \item $0.2 \; M_\odot \leq m \leq 0.9  \; M_\odot$: \textsf{PARSEC} isochrones \citep{bressan_parsec_2012,chen_improving_2014}. 
\end{itemize}

For hosts on the low-mass end, with $0.002 \; M_\odot \leq m < 0.2 \; M_\odot$, we perform a grid-search using the mass-appropriate evolutionary models to find spectra that have the closest match to the corresponding $\log g$ and effective temperature. We renormalize our model flux for each star to the Rosette Nebula distance (1400\,pc). Then, we convolve the flux with the F146 and F213 transmission curves \citep{cromey_science_2023} and derive the host's AB magnitude \citep{oke_secondary_1983}. We present the convolution of a model \citep{marley_sonora_2021} model atmosphere and the F146 bandpass in Figure~\ref{fig:transmission}. 

For hosts with higher masses, with $0.2 \; M_\odot \leq m < 1.5 \; M_\odot$, we instead use synthetic photometry derived from our solar-metallicity, 10\,Myr \textsf{PARSEC} isochrone (as shown in Figure~\ref{fig:isochrone}) for the Roman WFI bandpasses, typically given in Vega magnitudes. To convert this photometry to the AB magnitude system, we use \textsf{pysynphot} \citep{stsci_development_team_pysynphot_2013} to account for the correction factor. 

Lastly, we employ the Roman Exposure Time Calculator via the \textsf{Pandeia} engine \citep{pontoppidan_pandeia_2016} to estimate S/N for each host from our computed AB magnitudes. 
We report S/N for a 1\,hr long integration, implemented as 229 coadded exposures at an 18.97\,s cadence. With readout overheads, the elapsed time slightly exceeds 1\,hr. 
We adopt 1\,hr as a representative transit duration to benchmark single-transit detectability, acknowledging that actual durations vary and full campaigns will extend over multiple days.

In Table~\ref{tab:mags}, we present the resulting host AB magnitudes and S/N ratios for FFPs, BDs and stars between $0.002 \; M_\odot < m < 0.9 \; M_\odot$  (the F146 band saturation limit). The corresponding FFP S/N values indicate that such objects would be detectable down to ${\sim} 2\:M_{\rm J}$ over the two-week baseline. We note that these calculations do not include the effects from dust extinction. 
While the survey is sensitive to FFPs, the low S/N ratios per hour  ($5–20$) make transit detections improbable, except in rare cases of near–equal-size binaries on very short orbits.
However, the occurrence rate of such systems is expected to be rare \citep{canup_common_2006}. 
While the survey's sensitivity limits provide context for the detection threshold of low-mass objects, the primary goal of this study is to assess the detectability of transiting exoplanets around stars. 
Therefore, our filter selection is informed by the stellar-mass sources with markedly higher S/N values.
Between the F146 and F213 filters, the F146 filter is the optimal choice for maximizing sensitivity and yielding the most robust transit detection constraints for sources in the Rosette Nebula. 

\begin{deluxetable}{ccccccc}
\tabletypesize{\footnotesize}
\tablecaption{Host mass and radii at 10\,Myr. Source AB magnitudes are given in the F213 and F146 Roman filters, as well as the accompanying S/N value for target observations (distance of 1400\,pc is assumed). Magnitudes are provided without the inclusion of dust extinction.}
\label{tab:mags}
\tablehead{
\multicolumn{2}{c}{Host Mass} &
\colhead{Radius} &
\multicolumn{2}{c}{F213} &
\multicolumn{2}{c}{F146} \\
\colhead{($M_\odot$)} & \colhead{($M_\mathrm{J}$)} &
\colhead{($R_\odot$)} &
\colhead{(mag$_\mathrm{AB}$)} & \colhead{S/N (1-hr)} &
\colhead{(mag$_\mathrm{AB}$)} & \colhead{S/N (1-hr)}
}
\startdata
\multicolumn{7}{l}{\textit{Free-Floating Planets}}\\
\hline
0.002 & 2.1 & 0.136 & 29.53 & 0.04 & 28.64 & 0.7 \\
0.004 & 4.2 & 0.140 & 26.94 & 0.5 & 26.54 & 5 \\
0.006 & 6.3 & 0.145 & 25.34 & 2 & 25.15 & 16 \\
0.008 & 8.4 & 0.149 & 24.49 & 4 & 24.41 & 32 \\
0.010  & 10.5 & 0.153 & 23.71 &   9 & 23.79 &   56 \\
0.011  & 11.5 & 0.156 & 23.35 &  13 & 23.51 &   72 \\
0.012  & 12.6 & 0.161 & 23.01 &  18 & 23.22 &   92 \\
0.013  & 13.6 & 0.168 & 22.67 &  24 & 22.91 &  121 \\
\hline
\multicolumn{7}{l}{\textit{Brown Dwarfs}}\\
\hline
0.014  & 15   & 0.178 & 22.32 &  33 & 22.57 &  163 \\
0.015  & 16   & 0.190 & 21.81 &  53 & 22.03 &  253 \\
0.016  & 17   & 0.202 & 21.68 &  59 & 21.90 &  282 \\
0.017  & 18   & 0.213 & 21.41 &  75 & 21.60 &  358 \\
0.020  & 21   & 0.244 & 21.06 & 102 & 21.22 &  478 \\
0.030  & 31   & 0.283 & 20.43 & 175 & 20.49 &  807 \\
0.040  & 42   & 0.292 & 20.23 & 208 & 20.28 &  924 \\
0.050  & 52   & 0.315 & 20.07 & 238 & 20.12 & 1022 \\
0.060  & 63   & 0.335 & 19.82 & 291 & 19.83 & 1218 \\
0.070  & 73   & 0.358 & 19.68 & 326 & 19.69 & 1330 \\
\hline
\multicolumn{7}{l}{\textit{Stars}}\\
\hline
0.075  & 79   & 0.369 & 19.50 & 374 & 19.49 & 1493 \\
0.080  & 84   & 0.383 & 19.42 & 398 & 19.41 & 1564 \\
0.090  & 94   & 0.405 & 19.30 & 437 & 19.29 & 1675 \\
0.100  & 105  & 0.416 & 19.14 & 494 & 19.11 & 1849 \\
0.110  & 115  & 0.435 & 19.04 & 531 & 19.02 & 1950 \\
0.130  & 136  & 0.475 & 18.85 & 611 & 18.82 & 2162 \\
0.150  & 157  & 0.509 & 18.58 & 730 & 18.54 & 2504 \\
0.170  & 178  & 0.544 & 18.44 & 805 & 18.40 & 2695 \\
0.200  & 210  & 0.680 & 18.40 & 828 & 18.40 & 2691 \\
0.300  & 314  & 0.805 & 17.79 & 1229 & 17.75 & 3731 \\
0.400  & 419  & 0.895 & 17.37 & 1568 & 17.32 & 4579 \\
0.500  & 524  & 0.966 & 17.05 & 1884 & 16.98 & 3680 \\
0.600  & 629  & 1.020 & 16.78 & 2175 & 16.71 & 3919 \\
0.700  & 733  & 1.060 & 16.53 & 2490 & 16.45 & 4429 \\
0.800  & 838  & 1.020 & 16.22 & 2909 & 16.07 & 5260 \\
0.900 & 943 & 1.090 & 16.02 & 3224 & 15.84 & 5840\\
\enddata
\end{deluxetable}

\subsection{Transit Detection} \label{subsec: transit_detection}

While we perform detailed simulations to predict transit yields in Section~\ref{sec: simulation}, we first perform a simple calculation to estimate the exoplanet detection limit in the case of a single transit for the Rosette Nebula.
To this aim, we combine our synthetic photometry and S/N calculations with analytic models.
Consistent with prior investigations of ultra-short-period planets \citep{sanchis-ojeda_study_2014,ofir_independent_2013,smith_k2-137_2018}, we assume an orbital period of ${\sim}$8 hours for the transiting companion.

We compute transit durations under the assumption of circular orbits and edge-on inclinations, using the equation given by \citet{winn_exoplanet_2010}: 
\begin{equation} \label{eqn:winn_transit}
    t_\mathrm{dur}= \frac{P}{\pi} \sin^{-1} \left[ \frac{R_\mathrm{h}}{a} \frac{\sqrt{(1+k)^2 - b^2}}{\sin{i} }\right]
\end{equation}
where $P$ is the orbital period, $R_\mathrm{h}$ is the radius of the host, $a$ is the semi-major axis, $k$ is the companion to host radius ratio, $b$ is the orbital impact parameter, and $i$ is the inclination. For each host mass (and corresponding radius), we estimated the median transit duration from the system's ${\sim}$\,Roche-limit up to 0.2\,AU to account for a transitional Roche region, rather than a strict cutoff \citep{leinhardt_tidal_2012}.

\begin{figure}[h]
    \centering
    \includegraphics[width=\linewidth]{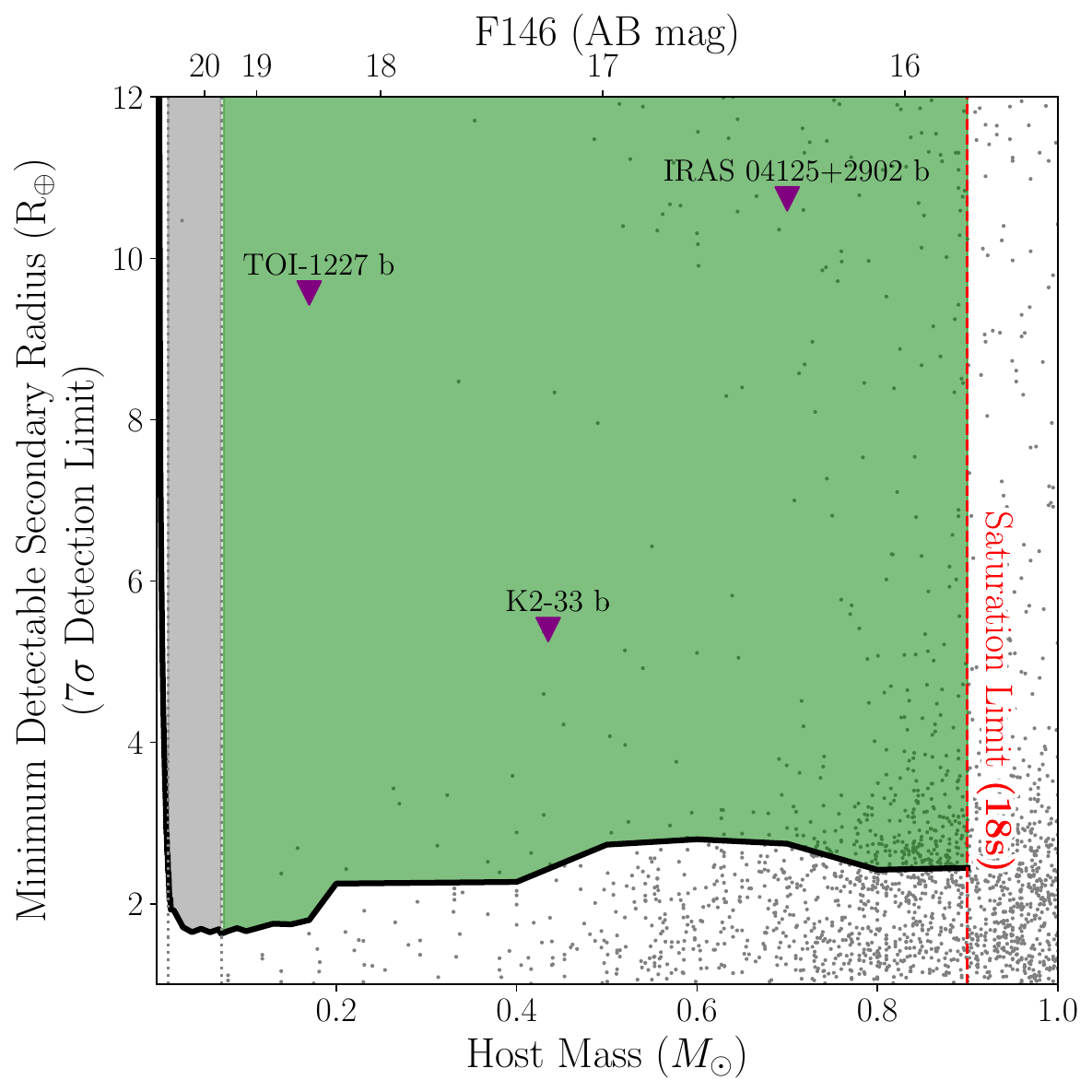}
    \caption{We show our $7\sigma$ detection limit on companion radius as a function of host mass using the black solid line. The green shaded region shows the range of detectable exoplanets around stars whereas the gray shaded shows the range of detectable exoplanets around brown dwarfs. The gray circles show all exoplanets from the NASA Exoplanet archive and the purple markers indicate transiting planets that are $< 20$\,Myr old in this parameter space.}
    \label{fig:f146detection}
\end{figure}

We then calculate the minimum detectable planet radius by requiring a 7$\sigma$ detection over an hour-long single transit. For our survey design in the F146 band, this corresponds to a sensitivity of planets ${\sim}$2\;$R_\oplus$ orbiting stars as massive as ${\sim}$0.9\;$M_\odot$, beyond which we approach our saturation limit. 

We find that the higher S/N achieved in the F146 filter enables the detection of transiting exoplanets orbiting late-type stars, where the planetary occurrence rates are the highest (small planets are $3.5\times$ more frequent around M dwarfs than F or G types; \citealt{mulders_stellar-mass-dependent_2015}), as well as up to solar mass hosts. This reaffirms the Rosette Nebula as a strong candidate for exoplanet discovery. Whereas the TEMPO survey is optimized for very-low mass hosts (FFPs and BDs), the distance to the Rosette Nebula extends the F146 saturation limit to more massive stars, allowing the survey to monitor a broader population of young stars. This is particularly significant because short-period planets are expected to be common in young systems \citep{rizzuto_zodiacal_2018, david_four_2019}, and their inflated radii \citep{fortney_planetary_2007, rogers_unveiling_2021} further enhance detectability. As shown in Figure~\ref{fig:f146detection}, only three transiting exoplanets younger than 20\,Myr have been identified \citep{barber_giant_2024,david_neptune-sized_2016,mann_tess_2022} around FGKM stars. The Rosette Nebula's young age provides a unique opportunity to extend exoplanet demographics to this scarcely-sampled age regime. 

\section{Transiting Exoplanet Yields with the Rosette Survey} \label{sec: simulation}

The previous calculation outlined the parameter space over which we are sensitive, but here we seek to estimate the number and demographics of young exoplanets we might detect. In order to determine a transit yield-rate for the Rosette Nebula, we develop a Monte Carlo injection-recovery framework that creates a sample population and models our detectability of transiting exoplanets in this system. By simulating a model population of one million hosts (500,000 FGK hosts and 500,000 M dwarfs), we determine a transit detection rate given our detectability of these companions over our temporal baseline of interest.

\subsection{Modeling Stellar Variability} \label{subsec: variability}

\begin{figure*}[ht!]
\centering
\includegraphics[width=0.9\textwidth]{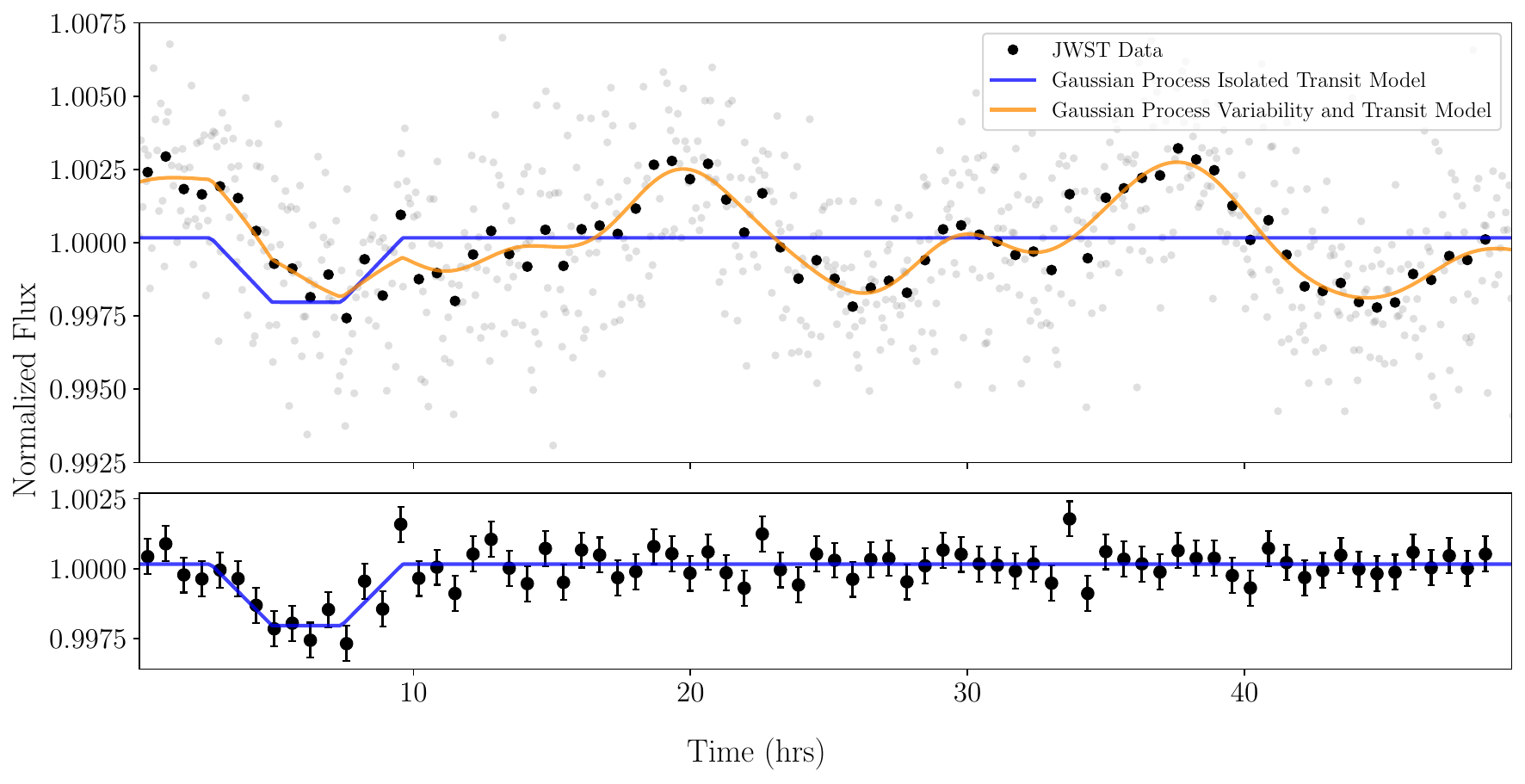}
\caption{We demonstrate that we can detect transits even in the presence of stellar variability, which is expected to be ubiquitous among young stars. \textbf{Top panel:} Light curve simulated in the Roman F146 filter (based on spectrally binned JWST/NIRSpec data for a H = 15.8\,mag, T6.5 methane-rich BD; \citealt{burgasser_discovery_1999}) with an injected transit. The light curve is fit with a Gaussian Process + transit model, which accurately recovers the injected transit at the injected location. \textbf{Bottom panel:} Detrended light curve showing the recovered transit detection. This injected transit (915\,ppm) lies at our $7\sigma$ detection threshold in the photon-noise-limited case discussed previously, demonstrating that this threshold is an appropriate measure of the transits we can detect even in the presence of stellar variability and instrumental noise.
}
\label{fig:simulated_lc}
\end{figure*}

Young stars are known to exhibit substantial variability \citep{barber_giant_2024}, generally making transit detection more difficult than for their older main-sequence counterparts. When calculating our transit detection sensitivity, we assumed that our capability would be photon-noise limited. However, stellar variability may also become a limiting factor in detection, as it is particularly pronounced at young stellar ages. We therefore adopt a conservative $7\sigma$ detection threshold---consistent with \citetalias{limbach_tempo_2023} and other transit surveys \citep{jenkins_impact_2002,borucki_characteristics_2011,burke_terrestrial_2015}---which provides both reliability and completeness, particularly in the presence of stellar variability in young stars. Here, we briefly demonstrate that we remain capable of detecting transits that meet our $7\sigma$ detection threshold, even in the presence of stellar variability.

We use archival James Webb Space Telescope (JWST) NIRSpec data of a variable star and construct a light curve by spectrally binning to the F146 Roman filter. This particular source exhibits a much faster rotation rate than expected for young stellar objects, so we stretched the light curve by a factor of ten, resulting in a rotation period of 17.7\,hr, typical for stars of this mass and age range \citep{herbst_rotation_2007}. The resulting simulated data are shown by the light-gray points in Figure~\ref{fig:simulated_lc}. The flux level and noise properties of this light curve are consistent with those expected for a 19.32\,$\mathrm{mag_{AB}}$ (or 0.09\,$M_\odot$) star in the Rosette Nebula.

We follow the methodology described in \citet{limbach_occurrence_2024}, using injection recovery with a Gaussian Process (GP) + transit framework to model variability and quantify detection completeness in time-series photometry. To illustrate that we can detect transits at the $7\sigma$ limit, we inject one of the transits identified in our simulations that was near this detection threshold. This transit has a depth of 915\,ppm and a duration of 4.29 hours. In the top panel of Figure~\ref{fig:simulated_lc}, we show the light curve with the injected transit, along with the fitted transit and variability model, whereas the bottom panel shows the detrended light curve with the successfully recovered transit. As depicted, our transit-injection recovery pipeline successfully recovers this signal, indicating that even in the presence of realistic stellar variability from space-based photometry, $7\sigma$ detections remain robust. While this demonstration considers only a single-transit event, given our proposed long temporal baseline, real observations will capture multiple transits for most short-period planets. Hence, for these systems, we note that our adopted $7\sigma$ threshold is a conservative lower bound on detectability. 

All the \textit{JWST} data used in this work can be found on MAST: \dataset[10.17909/ettj-kz68]{http://dx.doi.org/10.17909/ettj-kz68}. 

Young stars also exhibit prominent variability driven by circumstellar extinction events (dippers) and magnetic activity (flares), but these phenomena can often be distinguished from planetary transits based on geometric constraints and morphological differences. 

Dippers, which are the quasi-static or aperiodic dimming caused by occulting dust structures in the inner circumstellar disk \citep{cody_csi_2014,cody_many-faceted_2018}, typically last 0.5--2 days \citep[as also seen in the ${\sim}$10\,Myr Upper Scorpius region;][]{ansdell_young_2016} and exhibit irregular, asymmetric morphologies with variable depths between events. In contrast, planetary transits result in more symmetric dimming in the stars brightness with durations determined by Equation~\ref{eqn:winn_transit}. Combining this with Kepler's third law and assuming an edge-on, circular orbit with $k \ll 1$, we have a relation between orbital period and transit duration.

Using the stellar mass-radius relation from Table~\ref{tab:mags}, a 0.1\,$M_\odot$ M dwarf with a 12\,hr transit corresponds to an orbital period of ${\sim}450$\,days. Hence, dippers, with their characteristic long durations, would require orbital periods of several hundred days. Planets with orbits such as these are temporally separated from our sample of recovered transits (Section~\ref{subsec: results}), which have durations of a few hours, allowing dippers to be distinguished.

Lastly, flares present distinct geometric and morphological signatures when compared to both dippers and transits. Stellar flares are asymmetric brightness increases with characteristic fast-rise (minutes) and exponential-decay (tens of minutes to hours) timescales \citep{davenport_kepler_2016}. This asymmetric temporal profile can be readily distinguished from the symmetric ingress and egress of planetary transits, where the timescales are determined by the planet's velocity across the stellar disk. In addition to traditional transit-fitting methods, we can use automated flare identification algorithms such as \textsf{stella} \citep{feinstein_stella_2020} that are convolutional neural networks trained on these morphological features to robustly identify and mask flare events prior to transit searches.

While our injection-recovery demonstration shows that transits remain detectable in the presence of typical stellar variability, we acknowledge that more extreme variability events could impact our detection completeness. For instance, a large-amplitude flare increasing stellar flux by ${\sim}$20\% would obscure a simultaneous 0.5\% planetary transit. However, recent analyses by  \citet{householder_sensitivity_2025} and \citet{wilson_deep_2025} demonstrate that (1) even with more extreme variability amplitudes than our demonstrated case, transit detection algorithms do not produce false positives, and (2) detection thresholds remained close to the photon-noise limit rather than being dominated by stellar activity. This suggests that while extreme variability events may impact our detection limit, the time-averaged detection capability over our 14--30 day baseline approaches the photon-noise-limited case without introducing systematic false positives. Nevertheless, we acknowledge that our predicted yields may be somewhat optimistic if the Rosette Nebula hosts a population of extremely active stars with persistent high-amplitude variability exceeding that observed in WISE J1049-5319 AB and WISE J085510.83-071442.5.

\subsection{Creating the Model Population} \label{subsec: model_creation}

We generate a forward model of the stellar-planet population into our observed plane and then apply the survey parameters described in Section~\ref{sec: survey_design}. For each star, we draw planetary systems from a spectral-type dependent occurrence rate model, map their age-agnostic radii to an age-dependent one, propagate line-of-sight dust to apparent magnitudes, and evaluate detectability with a stellar-variability noise model to determine transit search thresholds. We describe our procedure in further detail below. 

\subsubsection{Dust Extinction and Stellar Hosts} \label{subsubsection: hosts_and_dust}

We model line-of-sight dust using the \textsf{dustmaps} Python package \citep{green_dustmaps_2018} with the Schlegel, Finkbeiner, and Davis (SFD) dustmap (\citeyear{schlegel_maps_1998}, recalibrated by \citealt{schlafly_measuring_2011}) sampled on a $0.5^\circ \times 0.5^\circ$ grid centered on the Rosette field. We convert $E(B-V)$ estimates to $A_V$ assuming the mean Galactic total-to-selective extinction ratio, $R_V = 3.1$ \citep{cardelli_relationship_1989}. Our assumption is consistent with \citet{fernandes_probing_2012}, who found that most members in the core of the Rosette Nebula follow a normal extinction law, though localized deviations toward higher $R_V$ are possible in SFRs (adopting $R_V = 4.0$ instead would only increase the mean $A_{\rm F146}$ by ${\sim}0.4$ mag). We then convert $A_V$ the to F146 bandpass extinction using the \citet{fitzpatrick_correcting_1999} extinction law evaluated at the filter's effective wavelength. We find that the Rosette Nebula has a mean ${A}_\mathrm{F146} = 1.25$\,mag with a dispersion of 0.43\,mag and a range of 1.784\,mag. We randomly sample stars from Table~\ref{tab:mags} by spectral type, assigning extinction from a Gaussian centered at the field mean with $\sigma = 0.43$. 

\subsubsection{Planetary Occurrence Rates} \label{subsubsec: occurrence_rates}

Early planetary occurrence rate studies from the \textit{Kepler} mission have shown that both the size and orbital period distribution of planets vary systematically with spectral type, with traditional core-accretion models also predicting the outward movement of the snow line with stellar mass and luminosity \citep{kennedy_planet_2008,laughlin_core_2004,mordasini_characterization_2012}. For M dwarfs, planets are not only more numerous but occur at shorter orbital periods, often forming compact multi-planet systems with periods $\lessapprox 50$\,days \citep{dressing_occurrence_2015, mulders_stellar-mass-dependent_2015,hardegree-ullman_kepler_2019}. In contrast, for planets orbiting FGK stars, the occurrence distribution shifts towards longer orbital periods, with a relative deficit of very short-period small planets and a rising frequency of giant planets at periods of hundreds of days \citep{howard_planet_2012,kunimoto_comparing_2020}.

Recent modeling work by \citet{hsu_occurrence_2019,hsu_occurrence_2020} account for these differences in planetary occurrence rates by using an Approximate Bayesian Framework (ABC) to infer occurrence distributions in the radius-period space from \textit{Kepler} data. ABC is particularly well-suited for exoplanet demographics because it accommodates detection biases and incompleteness of transit surveys, while still allowing flexible parametric forms for the occurrence likelihood surface. In particular, \citet{hsu_occurrence_2019} provided occurrence rate maps for FGK stars, whereas \citet{hsu_occurrence_2020} extended this framework to M dwarfs, exploring both uniform and Dirichlet (a multivariate generalization of the $\beta$-distribution) priors. In this work, we adopt the FGK occurrence rates from these studies and employ the Dirichlet prior formulation for M dwarfs. Our choice of the Dirichlet prior is motivated by its ability to generate more realistic and physically consistent occurrence rate samples rather than a simple uniform prior   \citep{komaki_asymptotically_2012,gelman_bayesian_2013}. 

In bins with an asymmetric distribution where only the 83rd-percentile upper limits are reported, we instead construct a continuous occurrence surface using cubic radial basis function interpolation, a well-established method for smooth multi-dimensional approximation \citep[e.g.,][]{powell_theory_1992,buhmann_radial_2003}. Where the interpolated rate exceeds the reported upper limit, we adopt the latter, ensuring consistency with the statistical constraints while maintaining continuity across the radius-period space. 

Within this framework, each simulated star is assigned planetary companions drawn with empirically constrained occurrence distributions. Orbital inclinations (in particular, $\cos{i}$) for each system are then sampled from a uniform distribution \citep{winn_exoplanet_2010}, consistent with the co-planarity in observed orientations of multi-planet systems \citep{fang_architecture_2012,fabrycky_architecture_2014}. Together, these steps produce a realistic synthetic planet population that underpins our injection-recovery analysis of the Rosette Nebula. 

\subsubsection{Radii Inflation} \label{subsubsec: radii_inflation}

Planetary radii are not static with age but contract as planets cool and lose residual heat from formation. Evolutionary models predict that young giant planets can have radii inflated by tens of percent compared to their older counterparts, with the degree of inflation depending on mass, composition, and formation pathways \citep{baraffe_evolutionary_2003,fortney_planetary_2007,spiegel_spectral_2012}. Observational evidence reaffirms this idea: several young planetary systems host transiting planets with radii larger than expected for their masses and orbital separations (K2-33b and K2-25b, \citealt{mann_zodiacal_2017}; the multi-planet system V1298 Tau, \citealt{david_four_2019}). 

To account for these effects in our model population, we adopt evolutionary cooling models from \citet{linder_evolutionary_2019}, which provide radius-age scaling relations across a wide range of planetary masses. For each planet drawn from our occurrence rate distribution described in Section~\ref{subsubsec: occurrence_rates}, we first determine age-agnostic mass-radius estimates from \textsf{Forecaster} \citep{chen_forecaster_2017,chen_probabilistic_2016}, which is a Bayesian forecasting model for objects covering nine orders-of-magnitude in mass. We then use these values to choose a mass-appropriate model and determine the scale factor given the planetary radius at 10\,Myr. We present these scaling factors in Table~\ref{tab:scaling_factors}.  

\begin{deluxetable}{cc}
\tablecaption{We present age-dependent (10\,Myr) radius scaling factors as a function of planetary radius $r$ relative to the final radius $R_{\oplus, f}$. \label{tab:scaling_factors}}
\tablewidth{\linewidth}
\tabletypesize{\normalsize}
\tablehead{
\colhead{Planetary Radius} & \colhead{Scaling Factor}
}
\startdata
$r < 2 \: R_{\oplus,f}$ & 1.45 \\
$2 \: R_{\oplus,f} \leq r < 5 \: R_{\oplus,f}$ & 1.51 \\
$5 \: R_{\oplus,f} \leq r < 8 \: R_{\oplus,f}$ & 1.40 \\
$8 \: R_{\oplus,f} \leq r < 9 \: R_{\oplus,f}$ & 1.35 \\
$9 \: R_{\oplus,f} \leq r < 16 \: R_{\oplus,f}$ & 1.20 \\
\enddata
\end{deluxetable}

\subsection{Generating Synthetic Lightcurves} \label{subsec: generating_lightcurve}

For each simulated host system, we generate transit models with the \textsf{batman} package \citep{kreidberg_batman_2015} at the survey cadence of 18\,s, assuming uniform stellar disks and circular orbits. The injected transits are multiplied onto unity baselines and combined for multi-planet systems. We then add Gaussian noise consistent with the expected precision of each host star (Section~\ref{subsec: synthetic_photometry}), ensuring that the synthetic light curves reproduce survey noise characteristics while preserving transit depths. These lightcurves form the input for our injection-recovery pipeline. 

\subsection{Recovering Transiting Exoplanets} \label{subsec: recovery}
To assess the detectability of injected planets, we process each synthetic lightcurve with the Box-Least-Squares (BLS) algorithm \citep{kovacs_box-fitting_2002,hartman_vartools_2016}, implemented within \textsf{Lightkurve} \citep{lightkurve_collaboration_lightkurve_2018}. For every host, we search across a grid of trial periods from 0.5-500\,days and a range of transit durations appropriate for companions on both short and long period orbits. The maximum BLS periodogram peak corresponds to an initial candidate period, reference time of mid-transit, and detection signal-to-noise ratio for a given companion. 

Given that some simulated systems host multiple planets, we apply an iterative masking algorithm to disentangle overlapping signals: after each detection above our adopted $7\sigma$ threshold, the corresponding transit window is masked and BLS is re-run on the residuals. This procedure allows us to recover successive transiting planets in multi-planet systems rather than picking out the strongest signal. This iterative BLS framework is widely used in other transit surveys and algorithms \citep{hippke_transit_2019,li_kepler_2019,tey_tess_2023}, ensuring a well-validated recovery procedure. For every simulated planet, we record whether its orbit produced a transiting geometry and its recovered BLS statistic. We identify this as a detection if this statistic exceeds the $7\sigma$ threshold during one of the iterative passes. This yields a a consistent mapping between the injected synthetic population and the subset that would be observationally recoverable in this survey. 

While our detection pipeline requires a 7$\sigma$ significance threshold, we do not mandate multiple observed transits within the survey baseline (see Section~\ref{subsec: results} for multiple-transit detections by host spectral type). This choice reflects a balance between mitigating false positive rates and scientific completeness. Requiring multiple transits would severely limit sensitivity to short-period planets and eliminate the longer-period systems that can be valuable for constraining migration timescales and orbital architectures at 10\,Myr.

Single transits have been proven viable for young planet discoveries and have traditionally served as a window into the population of long-period exoplanets \citep{wang_planet_2015,foreman-mackey_population_2016}. Recent confirmation of systems such as TOI-2180 b \citep{dalba_tess-keck_2022} and TOI-2010 b \citep{mann_giant_2023} demonstrate that a single high-SNR event can provide crucial information via its transit duration and morphology to enable targeted follow-up \citep[{{\sf Namaste};}][]{osborn_single_2016}. In the context of the Rosette Nebula, inclusion of single transits will be critical to capturing a larger diversity of the 10\,Myr planetary census, especially for those orbital periods that might exceed our observation window. 

Unlike \textit{TESS}, which monitors nearby stars amenable to ground-based confirmation, the Rosette Nebula's distance makes traditional follow up more challenging. \textit{JWST} provides a compelling path for validating and characterizing high-priority candidates. The young age and inflated radii of these planets result in enhanced atmospheric scale heights favorable for transmission spectroscopy. Consequently, faint NIR targets in the Rosette Nebula are within \textit{JWST}'s capabilities for both ephemeris refinement and atmospheric characterization using NIRSpec or NIRISS \citep{rustamkulov_early_2023,feinstein_early_2023}. Additionally, ground-based interferometry also offers a powerful validation pathway. 
While individual telescopes may lack Roman's spatial resolution, interferometric facilities can achieve milliarcsecond-resolutions to identify unresolved stellar companions that could mimic transit signals. 

\subsection{Background Eclipsing Binaries} \label{subsec: background_binaries}

We assess the probability of background eclipsing binaries (BEBs) being misidentified as planetary transits. Roman's WFI has a pixel scale of $0.11''$ \citep{akeson_wide_2019}, comparable to that of \textit{HST}. At the Rosette Nebula's distance of 1400 pc, a conservative photometric aperture with radius $r_{\rm ap} = 0.2''$ corresponds to a physical scale of ${\sim} 280$ AU, enabling high spatial resolution. Given that the Rosette Nebula has a peak stellar density of $\Sigma = 300$ stars pc$^{-2}$, this also corresponds to an angular stellar density of $\rho_* = 1.4 \times 10^{-2}$ stars per arcsecond$^{-2}$. Hence, the probability of a background star falling within our photometric aperture is, 
\begin{eqnarray}
    P_{\rm blend} & = \rho_* \times \pi r_{\rm ap}^2.
\end{eqnarray}

This indicates that only ${\sim} 0.08\%$ of our photometric apertures contain background stars. Of the small fraction of apertures containing background stars, the average occurrence rate of eclipsing binaries (EBs) across field populations is $f_{\rm EB} \approx 1.2\%$ \citep{prsa_kepler_2011}. Furthermore, only shallow grazing eclipses can mimic planetary transits, further reducing $P_{\rm BEB}$. This exceptionally low BEB contamination rate is enabled by a combination Roman's excellent spatial resolution and the Rosette Nebula's modest stellar density compared to denser environments such as globular clusters \citep[e.g.,][]{gilliland_lack_2000}. 

However, we note that additional false positive mitigation will require standard vetting procedures. BEBs have been commonly identified by (1) centroid analysis during the transit, which can reveal photocentric shifts for blended sources \citep{batalha_planetary_2013} and (2) secondary eclipse searches, as EBs produce secondary eclipses of comparable depth while planets produce much shallower (or undetectable) occultations. Our quoted yields therefore represent robust estimates of genuine planetary detections with minimal BEB contamination. 

\subsection{Results} \label{subsec: results}

\begin{deluxetable}{lccr}[h!]
\tablehead{\colhead{Host Type} &\colhead{Baseline} &\colhead{Transit Rate}
&\colhead{Detection Rate} \\
& (days) & & }
\tabletypesize{\scriptsize}
\tablewidth{\linewidth}
\tablecaption{Transit Yield Detection Rates for the Rosette Nebula \label{tab: yield_rate}}
\startdata
\textbf{M} & 14 & $2.6\% \pm 0.6 \%$ & $2.07\% \pm 0.5 \%$ \\
 & 30 & $2.57\% \pm 0.6\%$ & $2.36\% \pm 0.6 \%$ \\
\hline 
\textbf{FGK} & 14 & $1.31\% \pm 0.5 \%$ & $0.56\% \pm 0.3 \%$ \\
 & 30 & $1.32\% \pm 0.5 \%$ & $0.76\% \pm 0.3 \%$ \\
\enddata
\end{deluxetable}

\begin{figure}[h!]
    \centering
    \includegraphics[width=\linewidth]{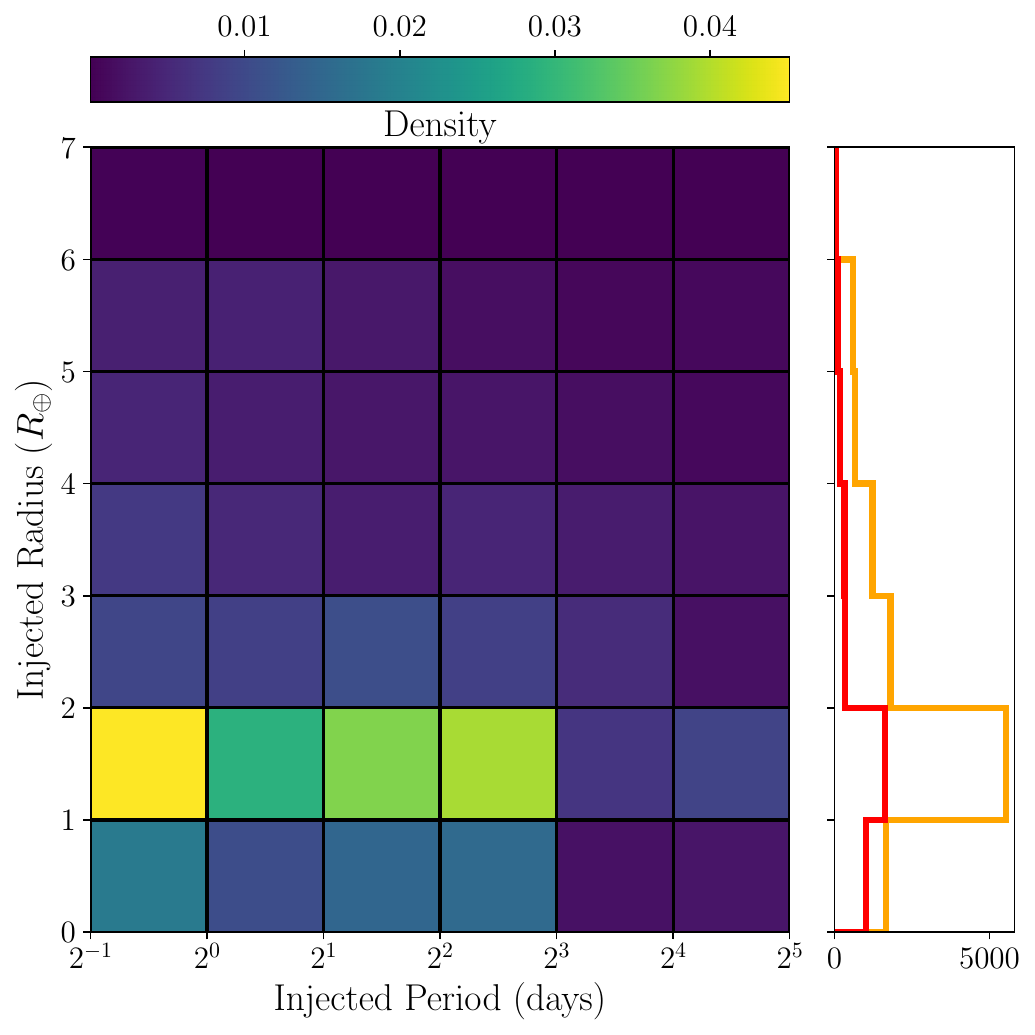}
    \caption{We present a 2D histogram with the distribution and density of transits recovered by host spectral type with one month of observations in radius-period space. We also show the marginal distributions of recovered planet radius for FGK stars (red) and M dwarfs (yellow). }
    \label{fig:recovery}
\end{figure}

\begin{figure*}[htbp]
    \centering
    \includegraphics[scale=0.7]{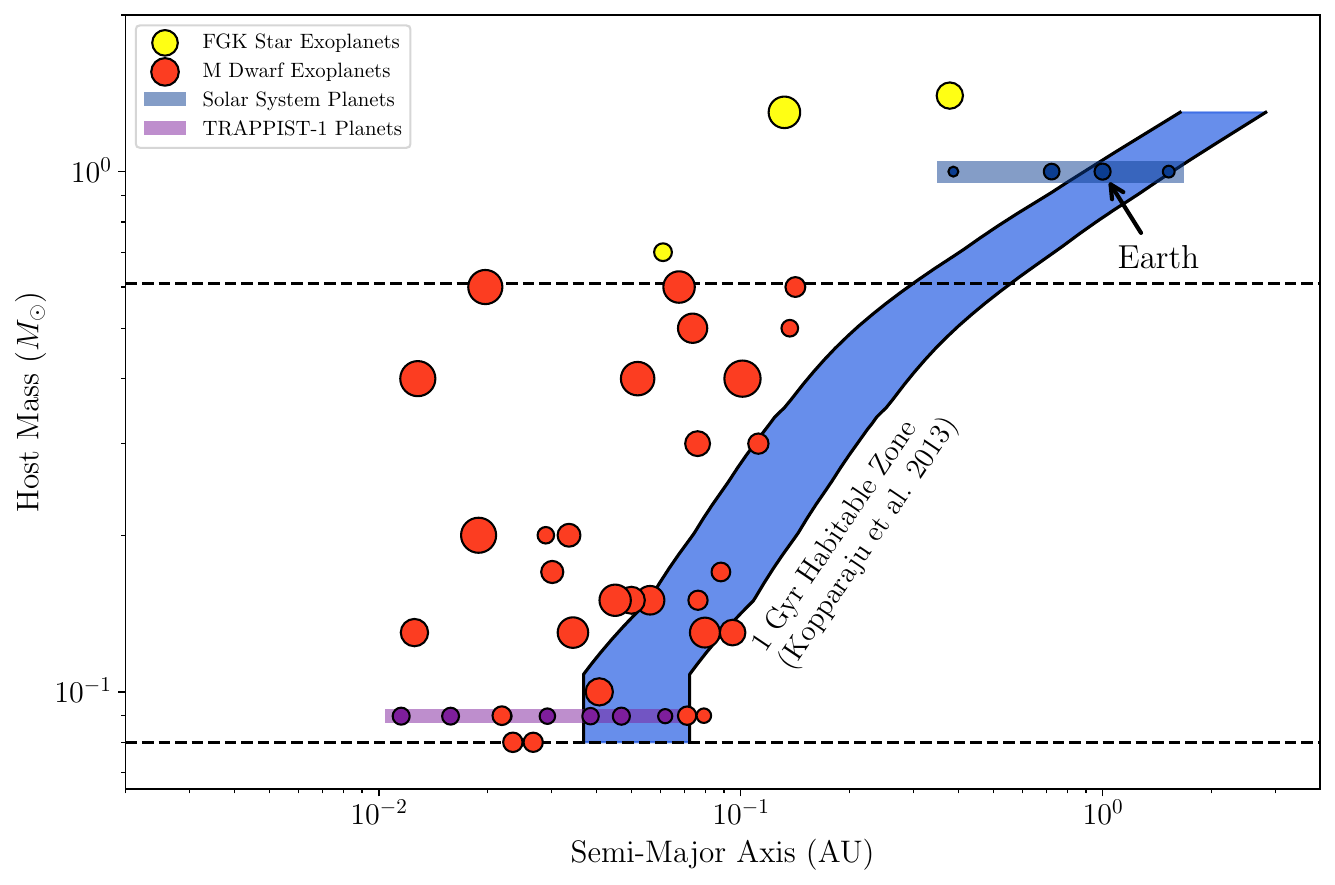}
    \caption{We show a simulated transiting population of exoplanets that Roman can detect in a month-long baseline. Red circles represent companions drawn from M dwarf occurrence rates whereas yellow circles represents companions drawn from FGK occurrence rates. We show the four innermost solar-system planets (dark blue circles) and the TRAPPIST-1 planets (purple circles). The sizes of each marker is scaled to the planet's radius. We also show the habitability zone at 1\,Gyr as estimated by \citet{kopparapu_habitable_2013} using \textsf{MIST} isochrones.}  
    \label{fig:simulation}
\end{figure*}

We run the full-injection recovery framework on our simulated population of one million host stars to quantify the expected yield of transiting exoplanets in the Rosette Nebula. We present our population transit and detection statistics for M dwarf and FGK hosts in Table~\ref{tab: yield_rate}. As expected, M dwarfs exhibit higher transit rates and detection rates than FGK stars, owing to their smaller host radii and compact planetary architectures. For a 14-day baseline, we find that 2.1\% of M dwarf planets are detectable, compared to only 0.6\% for FGK planets. Extending the observed baseline to 30 days increases our sensitivity to longer-period planets, particularly around FGK hosts, where the detection rate increases to 0.8\%. However, M dwarfs only show a smaller relative gain, potentially indicating that most detectable short-period planets are well-sampled in the two-week baseline. 

To convert these detection fractions to yield counts, we scale by the underlying stellar population we infer from the IMF (Section~\ref{subsec: imf}). We estimate $409$ young FGK stars, $1290$ young M dwarfs, and an additional $1125$ BDs (Table~\ref{tab:survey_overview}). Applying the detection rates from Table~\ref{tab: yield_rate}, we predict that a one-month baseline would a yield $3\pm1$ planets orbiting FGK hosts and $30^{+8}_{-7}$ planets orbiting M dwarfs. Likewise, we predict that a two-week baseline would yield $2\pm1$ planets orbiting FGK hosts and $27\pm7$ planets orbiting M dwarfs. 

Despite being able to monitor BDs, given the absence of well-constrained BD planet detections, we do not extrapolate our adopted occurrence rates down to this mass regime. Given our survey's capability of monitoring BDs at high S/N, detecting transiting companions around substellar hosts represents a novel and high-impact discovery space. Only a handful of BD-planet systems are known \citep[e.g.,][]{chauvin_giant_2004,bennett_low-mass_2008,han_microlensing_2013,jung_ogle-2017-blg-1522_2018}, and their origins remain ambiguous between planetary and binary formation pathways \citep{payne_potential_2007,bowler_imaging_2016}. Given that young BDs are infrared bright, this survey of the Rosette Nebula offers excellent sensitivity to detect transiting planets orbiting these substellar hosts, potentially helping establish one of their first statistical samples in a young coeval population. Thus, detecting even a single transiting planet around a BD would represent a fundamentally new contribution to exoplanet demographics at the substellar boundary. 

We show the overall distribution of recovered detections in period-radius space in Figure~\ref{fig:recovery}. The highest detection density occurs for planets with radii of 1-2\,$R_\oplus$ and periods less than ${\sim}8$\,days, consistent with the high geometric transit  probabilities and elevated occurrence rate of small planets at short orbits. We note the presence of a slight secondary excess for periods $\gtrapprox 32$\,days, which is driven by the coarser binning of the underlying occurrence-rate maps, where the our interpolation tends to overestimate frequencies and hence we cap occurrence rates at the 83rd-percentile upper limits. While the bulk of our detections arise from compact, short-period planets, the distribution also reveals a tail towards larger radii and longer periods. 

Our simulations also indicate that a majority of detections in a month-long baseline would multiple transits, with 64\% for M dwarf planets and 54\% for FGK star planets. These fractions reduce to 51\% and 39\% respectively for the two-week baseline, indicating that FGK planets benefit more from extended temporal coverage. 

Figure~\ref{fig:simulation} illustrates an example of a simulated transiting planet population we recover in the Rosette Nebula, with the yield counts consistent with a month-long baseline, where the sizes of each marker are scaled to their radii. The distribution emphasizes the dominance of super-Earths and sub-Neptunes, but also highlights the presence of larger detectable planets at wider separations, underscoring the richness of this survey's detectable population. We also overlay the 1\,Gyr habitable zone-track from \citet{kopparapu_habitable_2013}. We use Runaway Greenhouse and Maximum Greenhouse coefficients from their erratum-updated Table 3 combined with a 1\,Gyr, solar-metallically \textsf{MIST} \citep{dotter_mesa_2016,choi_mesa_2016,paxton_modules_2013,paxton_modules_2015} isochrone track to determine this zone. Given that most transiting planets at this age are interior to the habitable zone, this underscores Roman's potential to map out the evolutionary pathways of planets that may later enter habitable conditions. The Roman Space Telescope's yields in the Rosette Nebula may therefore represent the progenitor population of future habitable-zone planets, allowing us to to directly probe the initial conditions and evolutionary pathways that support the formation of life. 

\section{Summary} \label{sec: summary}

We investigate the scientific potential of a Nancy Grace Roman Space Telescope survey of the Rosette Nebula, a ${\sim}10$\,Myr star-forming region that provides a unique opportunity to study planets during their formative stages of evolution. Through modeling of the stellar population, photometric sensitivity analysis, and Monte Carlo injection-recovery simulations, we demonstrate that Roman can substantially expand the exoplanet census in a critical but poorly sampled age regime. 

Using \textit{Gaia} DR3 astrometry combined with clustering and kinematic searches, we identify 546 high-confidence candidate Rosette member stars within the Roman/WFI footprint. Extrapolating the initial mass function, we estimate that this survey will monitor 409 FGK-type stars, 1290 M dwarfs, and 1125 brown dwarfs at high signal-to-noise across the footprint. Observations in the F146 broadband filter optimize sensitivity for this mass range and achieve sufficient precision to detect sub-Neptune planets orbiting solar-type stars from a single-transit. 

Our simulations incorporate spectral-type-dependent occurrence rates, age-dependent radius inflation, and a conservative $7\sigma$ detection threshold. We predict the detection of $33\pm9$ transiting exoplanets in a month-long survey and $29\pm8$ exoplanets in a two-week survey, with approximately 90\% orbiting M dwarfs.
The extended baseline primarily improves sensitivity to longer-period planets orbiting FGK stars, while most M dwarf detections are captured within two weeks. 
These detections are dominated by of 1-2\,$R_\oplus$ super-Earths and sub-Neptunes with periods less than 8 days, representing an order-of-magnitude increase over the current sample of only three confirmed transiting planets younger than 20\,Myr around FGKM stars. 

This expanded sample would allow us to probe planetary demographics during late disk dissipation, directly addressing questions about radius evolution, migration timescales, and orbital stability. 
Some of these planets represent progenitors of future temperate worlds, allowing us to constrain the initial conditions and evolutionary pathways relevant to habitability. 
The inflated radii of young planets make them attractive candidates for atmospheric characterization with the James Webb Space Telescope, while ground-based facilities can provide mass measurements through radial velocity follow-up. 
Furthermore, long-term monitoring with the Vera Rubin Observatory can help constrain transit timing variations and stellar activity in the optical, and the future Habitable Worlds Observatory could directly image potential outer companions in these systems with the High-Resolution Imager. 

Moreover, the survey's sensitivity to brown dwarfs presents a unique and exhilarating science case. While we do not predict yields for this population due to uncertain planetary occurrence rates, detecting even a single planet transiting a young brown dwarf would provide crucial constraints on the lower limits of planet formation. 

The Rosette Nebula's youth, (sub)stellar richness, moderate extinction, and optimal distance create an exemplary environment for studying planets during a dynamic phase of their formation. With Roman's launch in late 2026, this survey presents a timely and transformative opportunity to address fundamental questions about planet formation that have remained observationally inaccessible.  

\begin{acknowledgements}

We are grateful to the anonymous reviewer for providing thoughtful comments that improved the scientific analyses presented in this paper. We gratefully acknowledge the generous support of NASA under Award No.~UGS25\_12-1 through the Wisconsin Space Grant Consortium. We also express gratitude to support from the \textit{Peter Livingston Scholars Program}, whose contributions to undergraduate research have played a key role in the development of this work.
Support for this research was provided by the Office of the Vice Chancellor for Research and Graduate Education at the University of Wisconsin--Madison with funding from the Wisconsin Alumni Research Foundation.
For our computational needs, our pipeline used \textsf{HTCondor}, a high-throughput computing cluster offered by the Center for High Throughput Computing (CHTC) at the University of Wisconsin-Madison. R.S.N. gratefully acknowledges Adam Distler for insightful discussions and comments during the preliminary development of this work's methodologies and analysis framework. 

This work has made use of data from the European Space Agency (ESA) mission \textit{Gaia} (\url{https://www.cosmos.esa.int/gaia}), processed by the \textit{Gaia} Data Processing and Analysis Consortium (DPAC; \url{https://www.cosmos.esa.int/web/gaia/dpac/ consortium}). Funding for the DPAC has been provided by national institutions, in particular the institutions participating in the \textit{Gaia} Multilateral Agreement. 

This work has also made use of the NASA Exoplanet Archive (\dataset[10.26133/NEA1]{http://dx.doi.org/10.26133/NEA1}), which is operated by the California Institute of Technology, under contract with the National Aeronautics and Space Administration under the Exoplanet Exploration Program.

This work was conducted at the University of Wisconsin-Madison, which is located on occupied ancestral land of the Ho-Chunk people, a place their nation has called Teejop since time immemorial. In an 1832 treaty, the Ho-Chunk were forced to cede this territory. The university was founded on and funded through this seized land; this legacy enabled the science presented here. 

JMV acknowledges support from the European Union through the Exo-PEA ERC project (grant number 101164652). Views and opinions expressed are however those of the author(s) only and do not necessarily reflect those of the European Union or the European Research Council Executive Agency. Neither the European Union nor the granting authority can be held responsible for them.
\end{acknowledgements}

\facilities{Gaia, 
James Webb Space Telescope, 
Mikulski Archive for Space Telescopes, 
NASA Exoplanet Archive}

\software{\textsf{Astroquery} \citep{ginsburg_astroquery_2019}, 
\textsf{Astropy} \citep{astropy_collaboration_astropy_2013,astropy_collaboration_astropy_2022,almendros-abad_spectroscopic_2023}, 
\textsf{batman} \citep{kreidberg_batman_2015},
\textsf{dustmaps} \citep{green_dustmaps_2018}, 
\textsf{HTCondor} \citep{thain_distributed_2005},
 \textsf{Lightcurve} \citep{lightkurve_collaboration_lightkurve_2018} 
\textsf{MIST} \citep{dotter_mesa_2016,choi_mesa_2016,paxton_modules_2011,paxton_modules_2013,paxton_modules_2015}, \textsf{NumPy} \citep{harris_array_2020}, \textsf{PAdova
and TRieste Stellar Evolution Code} \citep{bressan_parsec_2012,chen_improving_2014}, \textsf{pandas} \citep{mckinney_data_2010}, \textsf{pysynphot} \citep{stsci_development_team_pysynphot_2013},
\textsf{scikit-learn} \citep{pedregosa_scikit-learn_2011},
\textsf{SciPy} \citep{virtanen_scipy_2020}}

\clearpage

\begin{appendix} \label{sec:appendix}
We present normalized Kernel Density Estimates for the observed distributions of detected planets in our simulations.

\begin{figure}[h!]
    \centering
    \includegraphics[width=0.7\textwidth]{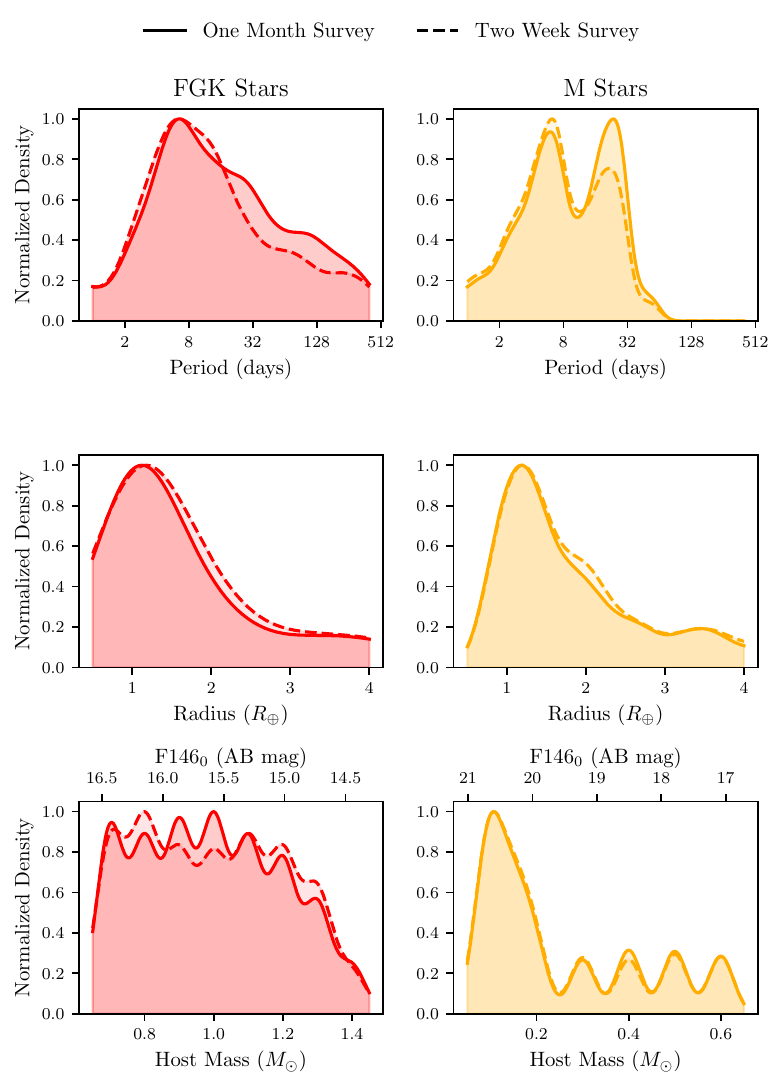}
    \caption{Normalized Kernel Density Estimates for detected planet candidates orbiting FGK stars (left) and M stars (right) comparing one-month (solid) and two-week (dashed) survey baselines. Rows show the distributions of orbital period, planet radius, and host stellar mass (with corresponding dust-corrected F146 magnitude on the secondary axis). }
    \label{fig:distributions}
\end{figure}
\end{appendix}

\clearpage

\bibliography{rosettetransityields}{}
\bibliographystyle{aasjournalv7}

\end{document}